# Proposition of an implementation framework enabling benchmarking of Holonic Manufacturing Systems

Olivier Cardin, Anne L'Anton

LUNAM Université, IUT de Nantes – Université de Nantes, LS2N UMR CNRS 6004
(Laboratoire des Sciences du Numérique de Nantes),
2 avenue du Prof. Jean Rouxel – 44475 Carquefou

(e-mail: olivier.cardin@ls2n.fr).

**Abstract** Performing an overview of the benchmarking initiatives oriented towards the performance evaluation of Holonic Manufacturing Systems shows that there are very few of them. However, a comparison between all the isolated emulation developments for benchmarking in literature was made, and showed that many common features could be extracted. Several deadlocks for a generic approach of these developments are also exhibited. A global architecture dedicated to a generic performance evaluation platform design is suggested. This architecture integrates a scenario manager, whose main specificities were detailed and justified. Those features are meant to both integrate the best practices encountered in literature and fulfil the missing aspects to respond to the problematics.

## 1 Introduction

Current research and developments in next generation manufacturing control systems, and specifically Holonic Manufacturing Systems, recently emphasized the maturity of the underlying concepts and methods [1]. In this context, next step is a dissemination of the concepts, primarily through a wide industrial acceptance of the related developments. However, those control architectures suffer from a lack of performance guarantee, as they are mainly based on emerging behaviour techniques, such as multi-agent systems or holonic paradigm, making the performance of the control system highly dependent on the context of execution of the experiment [2].

Traditional benchmarking activity consists in evaluating the response of the control system to a manufacturing situation with a predefined set of data. Several years ago, the operational research (OR) community has proposed several benchmarks to try and compare the algorithms solving static NP-hard optimization problems for production, among which [3] is one of the first. This approach is not fully satisfying for next generation control systems. Indeed, their major interest relies in their robustness and reconfiguration abilities, that requires to be evaluated online [4]. Therefore, a whole new evaluation framework including both the final control system plugged on an emulation/virtual representation of the manufacturing system, in one of the scenarios called High Level Virtual Commissioning expressed in [5]. A performance evaluation conceptual framework was developed for assessing the level of quality of a scheduling solution in terms of efficiency, robustness and flexibility [6], and defined several years ago the general architecture of an online benchmarking instance (Fig. 1) which exhibits perfectly the full decomposition between control system and emulation.

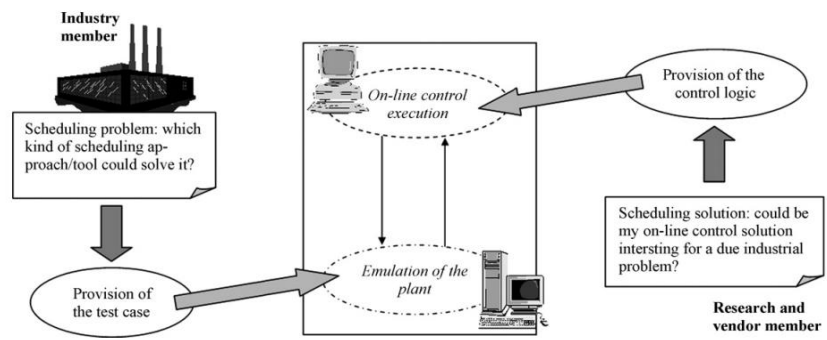

**Fig. 1. General framework of an online benchmarking instance** [6]

The main problematic the community is currently facing is the lack of details of this generic framework, making each application developed ad hoc with various functionalities and possibilities. The aim of this paper is to suggest an implementation framework of both the control system and the emulation model in order to standardize the development of such initiative and allow the application of various benchmarks.

Therefore, a comprehensive analysis of the existing benchmarks in literature is performed in second section in order to exhibit the requirements for the range of scenarios to incorporate in the framework. Then, a review of some of the existing emulation model developed in literature is proposed in third section in order to emphasize the convergence between each individual initiative. Finally, the resulting framework is presented in the fourth section.

## 2      Benchmarking Holonic Manufacturing Systems

Evaluating the performance of Holonic Manufacturing Systems is reputed to be a difficult task, as it requires a dynamic evaluation of the control system's response to predefined scenarios, as much as a prerequisite for industrial acceptance. As a matter of fact, numerous works in literature can be found that exhibit a performance evaluation, but generally on ad hoc scenarios fully customised for the dedi-cated application. Among those, Jovanovic et al. [7] studies the implementation of a holonic control system on a "green"-tyre-manufacturing system. The objective is to evaluate how the holonic control is able to eliminate the impact of machine breakdowns on productivity. To do so, two examples of scenario are chosen, and the comparison with so called classical control approaches exhibits a 4% increase of productivity.

Table 1 synthesizes, for each of the initiatives presented in this section, the type of scenarios encountered. It states for each of the benchmarks encountered in literature (in columns) the category of scenarios that are taken into account. These categories are:

- Dynamic reconfiguration, impacting the whole system, typically machine failures;
- Quality issues, impacting mainly the products, typically rejection or remanufacturing of products;
- Order management, impacting the control system, typically cancelled or high priority orders;
- Supply issue, impacting the control system, typically a shortage in components on a machine.

**Table 1.** Holonic Manufacturing Systems benchmarking scenarios categorization

| Category | Directly impacted element | Unstable conditions [7] | Manufacturing disturbance scenarios [8] | Experimental modalities [9] | Dynamic production system scenarios [4] |
|---|---|---|---|---|---|
| Dynamic reconfiguration | Shop floor resources | Example 1; Example 2 | Query 2; Query 4; Query 5; Query 6; Query 9; Query 10 | PD1; PD2 | #PS2; #PS3; #PS3; #PS7; #PS9; #PS10; #PS12 |
| Quality issues | Products | | | | #PS6; #PS11 |
| Order management | Control system | | Query 3; Query 7; Query 8 | BD1; BD2 | #PS1; #PS4; #PS13; #PS14; #PS15 |
| Supply issues | Control system | | | | #PS8 |

Bal and Hashemipour [8] suggest a virtual reality-based methodology for enhancing the design and implementation process of holonic control systems in manufacturing practice with the objective of implementing and disseminating holonic control into the small to medium size manufacturing enterprises. The case study is developed on a die-casting factory, *Sahin Metal*, in Istanbul, Turkey. The objective is to measure Throughputs, Lead Times and Resources utilization considering 10 different scenarios, called "Manufacturing disturbance scenarios". Those scenarios extend the range of considered cases in the following directions:

- The reconfiguration of the system also consider rapid insertion of new resources or modification of their capabilities;
- The information system is considered, with the management of the order and their dynamic evolution (rush orders for example – see Table 1).

Even if Table 1 exhibits a relative convergence of the considered scenarios, initiatives tried to define a full methodology to design the benchmarking experiment. Among those, Mönch [10] suggested the following scheme in order to construct the benchmark:

1. Determination of production control approaches used for comparison;
2. Determination and specification of the used performance measures;
3. Specification of the used performance assessment strategy;
4. Description of the hardware and software environment for the benchmark;
5. Description of different scenarios that should be simulated. This includes especially the description of designed experiments;
6. Simulation of the scenarios and discussion of the results.

Pannequin et al. [9] defined a benchmarking protocol, targeting HMS implementation projects. A component-based generic architecture is proposed with this protocol, enabling to model and compare various control architectures. The case study relies on an automotive-industry. Business oriented disturbances (BD) are considered (Order management) along with Process oriented disturbances (PD) that relate to Dynamic reconfiguration.

Finally, the Bench4Star initiative [4] is probably currently the most advanced benchmark for HMS in literature. As exhibited in Table 1, more scenarios are taken into account with Quality and Supply issues, which extend the range of the evaluations and make the scenarios closer to real manufacturing conditions.

## 3 Emulation of HMS

### 3.1 Development approaches

From the individual initiatives that were developed among the years in literature, an empiric approach in the development of emulation-based performance evaluation of HMS control was designed (Fig. 2) [8]. In the general approach (a), the emulation issue is mainly located in the bottom part. Part (b) of Fig. 2 represents with more details the development process of the virtual factory model. Two elements might be noticed. First, the scenarios are not mentioned, which implies the necessity to develop ad hoc models for each tested scenario. Second, the VR mod-el returns performance indicators for the analysis of the response of the control to the scenario.

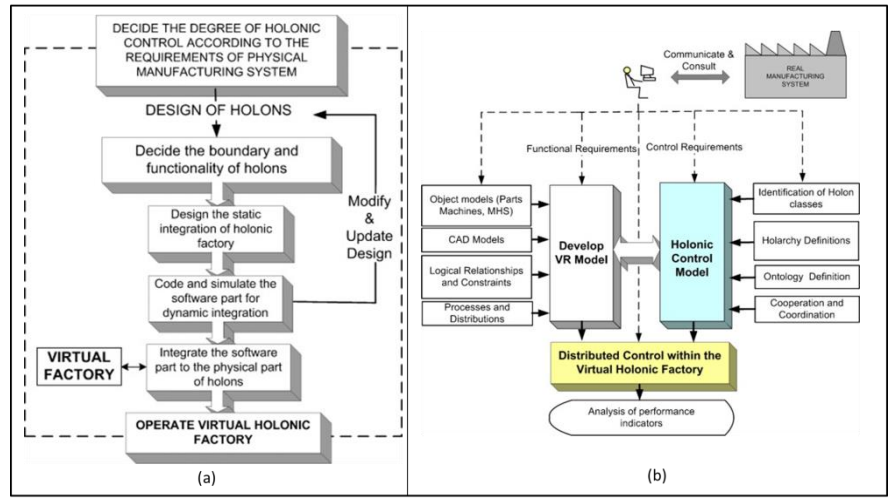

**Fig. 2. Design approach of emulation-based performance evaluation (a) and development detail of emulation architecture (b)** [8]

In the same way, a software architecture was suggested by [10] (Fig.3). It was designed for a full integration with C++-based control system and Java simulation tools, and a web-based access to allow the users building their own simulation models from scratch by specifying the simulation model in XML format. In this architecture, the coupling between the control algorithm and the emulation is loose, and the performance indicators are extracted both from the emulation and the control system.

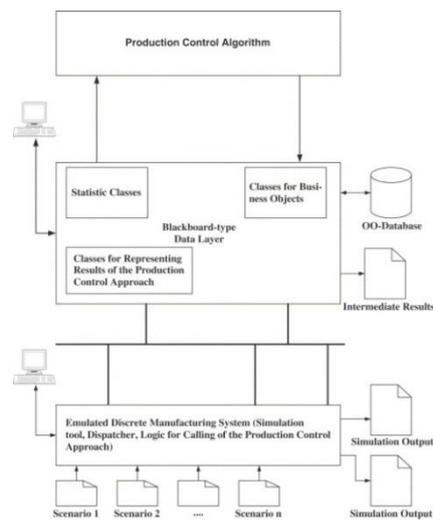

**Fig. 3.** Software architecture of emulation-based control [10]

The analysis of both these approaches exhibit several problematics:

- #P1: What coupling for a generic approach between holons and emulation?
- #P2: How to retrieve the performance indicators?
- #P3: Which integration of the scenarios in the architecture?

Next section intends to perform a literature review of the proposed developments and examine their response to these questions.

### 3.2  HMS emulation literature review

#### 3.2.1  Answers to #P1: coupling HMS/emulation

Several studies were lead on the genericity of the approach of emulation, such as [9] or [10] that were previously mentioned. Another interesting initiative was called Arezzo-FMS [5]. The idea was to develop a generic emulation model and generation methodology that was able to connect to the shop floor control, which is itself easily reconnected to the real shop floor (Fig.4).

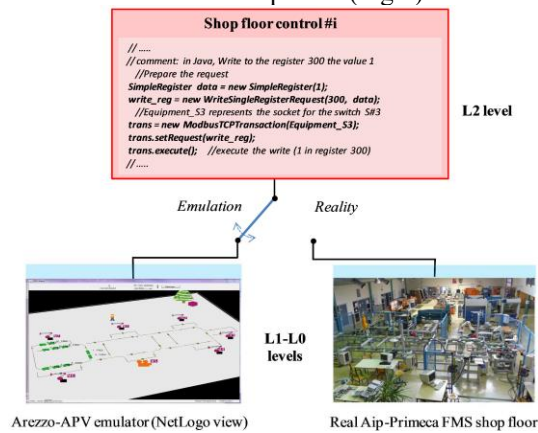

**Fig. 4.** Arezzo-FMS general scheme [5]

In this context, they introduced the concept of *Interface Layer (IL)*, which is one of the major development primitive, allowing the communication between holons and simulated objects the same way they do with real shop floor entities (Fig. 5). Examining the various studies that also exhibited the use of emulation for performance evaluation, this point is frequently dealt with (Fig. 6).

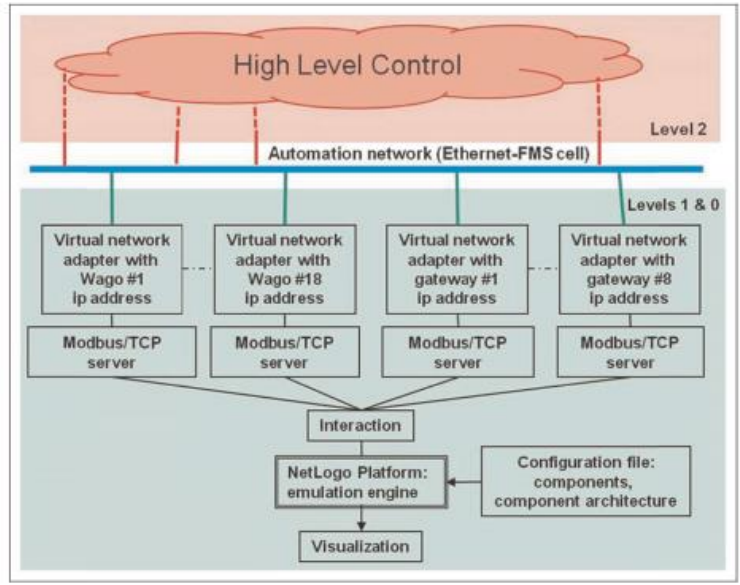

**Fig. 5. Arezzo-FMS general scheme [5]**

The three studies exhibited in Fig. 6 have various purposes: case a) is related to the holonic control in tyre manufacturing industry [7], case b) deals with the control of a flexible manufacturing system [11], whereas case c) intends to validate the behaviour of a holonic controller of modular conveyor systems [12]. They were developed in parallel without interaction, but show several common features. One of them is the presence of the IL at the interface between the virtual model and the real control to be tested. Identical conclusions can be drawn about cloud simulation platforms [13] or agent-based manufacturing systems [14] for example.

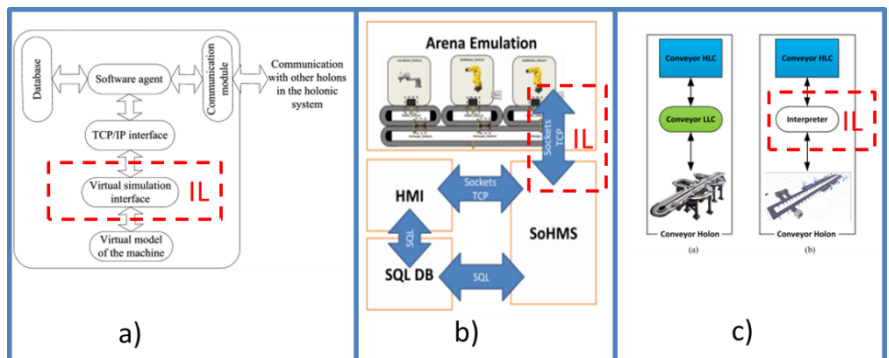

**Fig. 6. Some interface layers in literature [7], [11], [12]**

### 3.2.2 Answers to #P2: performance indicators

The question of the performance indicators (KPI) is dealt with in two main ways:

1. The emulation model is based on a discrete-event simulation tool, whose outputs are used as KPI, like in [8] for example;
2. The control system has its own KPI output module, which is used in an emulation study in the same way it would be in real cases, like in [7] for example.

However, most of the studies found in literature do not mention the way the KPI are gathered and calculated.

### 3.2.3 Answers to #P3: scenario integration

As far as the authors know about, this question was not deeply treated in a generic way in literature: all the developments were made ad hoc for unique performance evaluations. The only reference to such element can be found in [4] where the scenario is meant to be integrated in parallel with the initial data set of the control environment, but no indication is given on the way to achieve this integration in a dynamic environment.

This corresponds to the lack of predefined benchmark exhibited earlier in this article. Now that initiatives such as Bench4Star [4] rose, a generic scenario manager could probably be designed, enabling an easy coupling between Bench4Star and emulation initiatives. This constitutes the purpose of next section, which intends to design a global architecture integrating this scenario manager and specifying the expected characteristics.

## 4 Proposition of a generic implementation architecture

The scope of this section is to define the general architecture and prerequisites for the most valuable response to the problematics expressed before. Fig. 7 introduces the global architecture. It is based on a generic emulation-based architecture (left side of the figure) extracted from the previous analysis of literature. Considering all the works published, the following elements can be defined:

- Emulation model: simulation-based dynamic model of the real system;
- Control system: Holonic based control system to be evaluated. The human-machine interface was not represented apart here, however it could be;
- Production database: for orders management and relationship with tools such as ERP for example;
- Interface layer: as previously discussed, this layer intends to ease the switch between emulation-based evaluation and control of the real system.

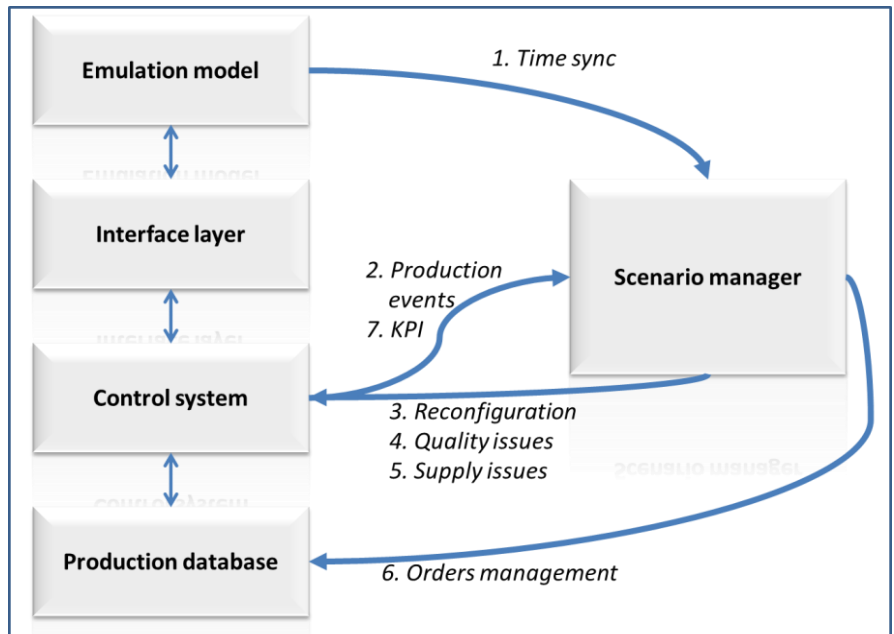

**Fig. 7. Integration of a scenario manager in an emulation based control architecture**

This last element is the main answer of the architecture to #P1 concerning the integration between holons and emulation. This interface layer shall be standardized in order to be implemented easier. The evolution of control systems, slowly migrating to the cloud [13], makes the problem of interoperability more and more important, and goes therefore in the right direction for this purpose. The proposition of using web services-like architectures, designing Service-oriented HMS [15], is probably a first step towards this objective.

Considering #P2 and the problem of performance indicators, both the options that are discussed in literature do not show on our point of view a good adequacy with the objectives of a generic approach for performance evaluation.

The use of the emulation model simulation outputs to calculate the KPI is very interesting for the utilization of the machines for example, but seems simplistic on a general point of view, as it prevents from getting KPI about the order management system for example, or about the behaviours of holons (decision making delay for example). Furthermore, one of our objectives for the emulation model is to be as lean as possible, so that it can be used in all scenarios without model modification. This is totally impossible with the use of the model for KPI calculation.

Another direction is to design the control system to be able to compute its own KPI. This is a very efficient solution, as this element of the architecture is aware of all the events that can perturb the performances of the overall system. However, it does not seem relevant to modify the design of the control system for emulation purposes: it would be better to use the full control system without modifications. Moreover, the variety of the studied scenarios and the expected associated KPI makes it a huge patch to integrate in the software that might modify the behaviour of the control system.

Therefore, the proposition here is to gather data (label 2. Of Fig. 7) or direct KPI (label 7.) from the control system and externalize the calculation of the KPI in another element of the architecture. As the expected KPI vary between each tested scenario, this element needs to know about the running scenario and about the actual time of the system. Indeed, the time of the system is dictated by the emulation model, whereas the control system does not necessarily know about it. This element therefore also needs a connection to the emulation model for data gathering (label 1.).

This last proposition leads to the definition of a "Scenario Manager", able to modify the behaviour of the control system according to the chosen scenario (labels 3., 4., 5. and 6.). For those last features, the scenario manager needs to have an access to the control system in various forms, but all these interactions are probably meant to be at least created for the human-machine interaction. The only one might be quality issues, where the actual information comes from the shop floor in real time execution. In that case, it is the scenario manager that needs to endorse this role and handle most of the random data distributions.

Fig. 8 shows a sequence diagram expressing the behaviour of the scenario manager in the case of scenario #PS9 extracted from [4]. This scenario needs a reconfiguration of the system due to machine breakdown. The problem is that the date of the breakdown is determined dynamically considering the departure date of the first shuttle from this machine. Therefore, the scenario manager needs production events to know when it needs to reconfigure the control system to take into account the breakdown of M2.

## 5     Conclusion

The objective of this paper was first to provide an extensive overview of the benchmarking initiatives oriented towards the performance evaluation of Holonic Manufacturing Systems. Then, a comparison between all the isolated emulation developments for benchmarking was made, and common features and main problematics were exhibited.

Finally, a global architecture dedicated to a generic performance evaluation platform design was suggested. This architecture integrates a scenario manager, whose main specificities were detailed and justified. Those features are meant to both integrate the best practices encountered in literature and fulfil the missing aspects to respond to the problematics. Basically, the idea is to develop a piece of software integrating a priori all the scenarios of literature benchmarks, with standardized interfaces and which would be able to modify in real time the behaviour of the system (triggering breakdowns, order management, etc.) and generate adequate performance indicators at the end of the scenarios runs. We believe this is the elementary brick missing to be really efficient in performance evaluation, but also a very difficult brick to develop in a generic way.

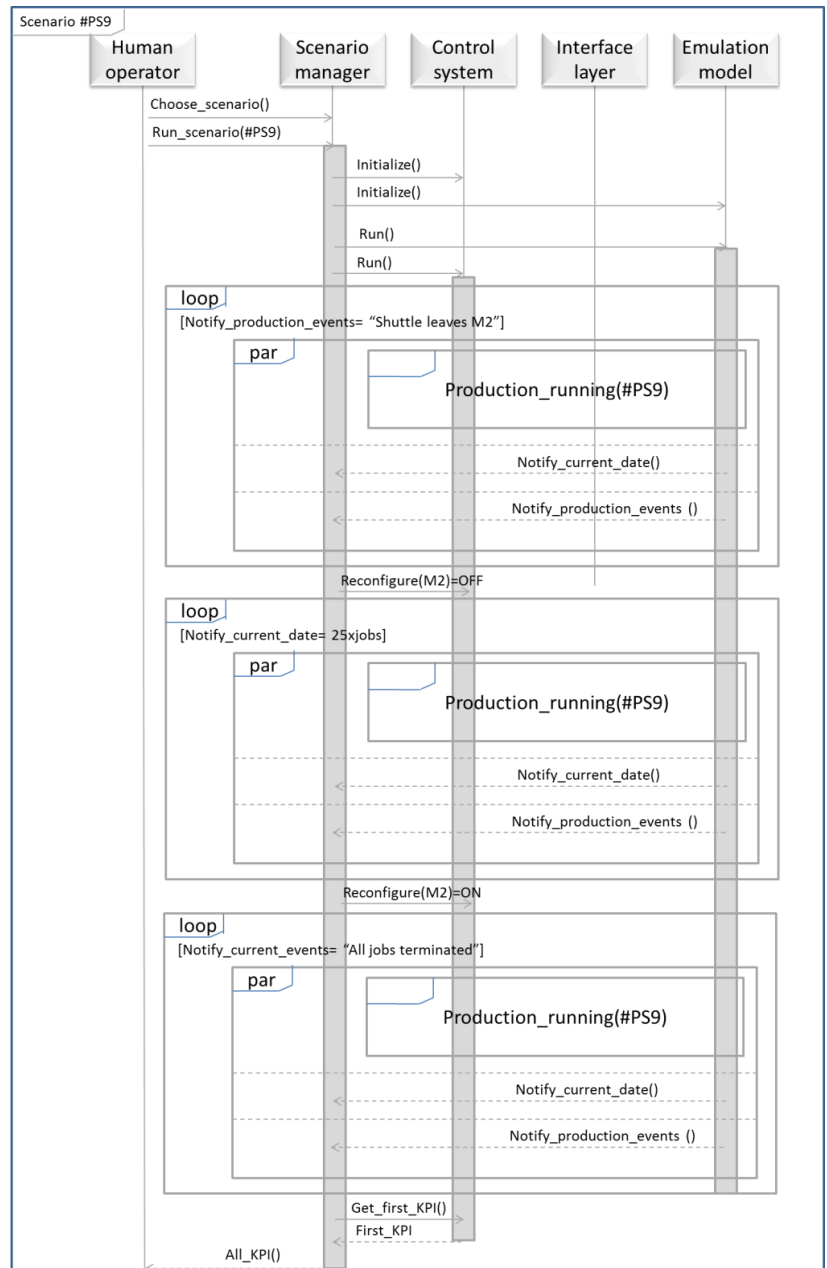

**Fig. 8.** Sequence diagram of #PS9 scenario integration

The main objective now is to foster a globalization of these considerations among the main actors of the domain in order to try and develop a scenario manager able to connect to most of the control systems developed in parallel in the community.